\documentclass[twocolumn,preprintnumbers,amsmath,amssymb]{revtex4}
\usepackage{graphicx}
\usepackage{dcolumn}
\usepackage{bm}

\begin{document}
\title{Thermomechanical properties of a single hexagonal
boron nitride sheet}

\author{Sandeep Kumar Singh$^1$, M. Neek-Amal$^{1,2}$\footnote{corresponding author:
neekamal@srttu.edu}, S. Costamagna$^{1,3}$, F. M. Peeters $^{1}$}
\affiliation{$^1$Department of Physics, University of Antwerp,
Groenenborgerlaan 171, B-2020 Antwerpen, Belgium\\ $^2$Department of
Physics, Shahid Rajaee University, Lavizan, Tehran 16785-136,
Iran\\$^3$Facultad de Ciencias Exactas Ingenier\'ia y Agrimensura,
Universidad Nacional de Rosario and Instituto de F\'isica Rosario,
Bv. 27 de Febrero 210 bis, 2000 Rosario, Argentina.}

\date{\today}

\begin{abstract}
Using atomistic simulations  we investigate the thermodynamical
properties of a single atomic layer of hexagonal boron nitride
(h-BN). The thermal induced ripples, heat capacity,  and thermal
lattice expansion of large scale h-BN sheets are determined and
compared to those found for graphene (GE) for temperatures up to
1000 K. By analyzing the mean square height fluctuations $\langle
h^2\rangle$ and the height-height correlation function $H(q)$ we
found that the h-BN sheet is a less stiff material as compared to
graphene. The bending rigidity of h-BN: i) is about $16\%$ smaller
than the one of GE at room temperature (300 K), and ii) increases
with temperature as in GE. The difference in stiffness between h-BN
and GE results in unequal responses to external uniaxial and shear
stress and different buckling transitions. In contrast to a GE
sheet, the buckling transition of a h-BN sheet depends strongly on
the direction of the applied compression. The molar heat capacity,
thermal expansion coefficient and the Gruneisen parameter are
estimated to be 25.2 J\,mol$^{-1}$\,K$^{-1}$,
7.2$\times$10$^{-6}$K$^{-1}$ and 0.89, respectively.

\end{abstract}

\maketitle

\section{Introduction}

A single layer of hexagonal boron-nitride (h-BN) is a wide
gap insulator that is very  promising for opto-electronic
technologies~\cite{14,15}, tunnel devices and field-effect
transistors~\cite{naonoletter2012,APL2011,Beheshtian}.
According to the well known Mermin-Wagner theorem~\cite{mermin}, the
stability of any two dimensional crystal is only possible in the
presence of atomic corrugations which distort the perfect honey-comb
lattice and allow the atoms to explore the out-of-plane direction.
Experimental observations have found ripples in suspended sheets of
GE~\cite{3,kirilenko} and atomistic simulations suggest that the
strong bonds between the carbon atoms determine the features of
these ripples~\cite{4}. Understanding the behavior of the ripples is
important for many reasons~\cite{seba2012}. They affect the
electronic transport properties, e. g., in GE the ripples are
believed to be one of the dominant scattering sources which limits
the electron mobility~\cite{geim-kat,seba1}.

h-BN ribbons doped by carbon has recently been
proposed~\cite{nanolett2012,Beheshtian}. In addition, BN based
nanostructures are potential materials for thermal management
applications~\cite{Ouyang, Savic1, Stewart, Chang,Shen,Xiao} because
of their  high thermal conductivity and sensitivity to isotopic
substitution and etc. Therefore, the knowledge of the shape and the
temperature dependence of the intrinsic ripples is fundamental to
devise novel nano-devices based on this material.

\begin{figure}
\includegraphics[width=0.48\textwidth]{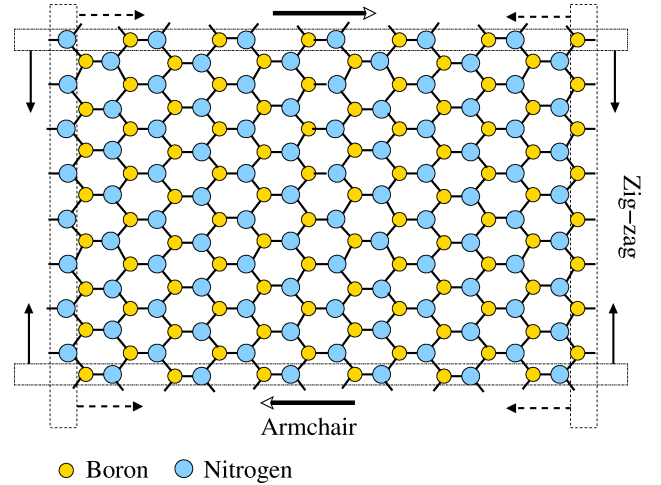}
\caption{ (Color online) Schematic view of the single h-BN sheet.
Smaller-yellow (bigger-blue) circles refer to the B (N) atoms.
The rectangles indicate the atoms that are fixed
during compression.
Dashed (straight) lines correspond to arm-chair (zig-zag) uniaxial compression in the direction
given by the arrows. Open arrows indicate the shear stress applied in the armchair direction.
}
\label{fig1}
\end{figure}

Both GE and h-BN sheets have a honeycomb
lattice structure, however the different
electronic properties and phonon band structure~\cite{ph,el1,el}
results in unequal morphologies and corrugations.
Transmission electron microscopy is widely used to resolve the
individual atoms in suspended h-BN sheets~\cite{PRB802009} where
ripples inherently exist.
First-principle calculations have been performed using small unit
cells, periodically replicated, which are unable to model long
wavelength corrugations which require thousands of
atoms~\cite{hasan} while the mechanical properties of a h-BN sheet can
be estimated by using DFT~\cite{boris}. Conversely, atomistic
simulations using molecular dynamics simulations (MD) enable to
study thermo-mechanical properties directly on large scale samples.
The modified Tersoff potential~\cite{Tersoff} (TP) (parameterized
originally for carbon and silicon) with various set of parameters
have shown to predict reasonable values for the thermo-mechanical
properties of the h-BN sheet. In the pioneer work by Albe~\emph{et
al} re-parameterized TP was used to study the impact of energetic
boron and nitrogen atoms on a cubic-BN target~\cite{Albe}.
Some other groups have also parameterized TP using various
experimental data, e.g. Sekkal \emph{et al}~\cite{Sekkal} treated
h-BN as a one-component system, using the same  potential parameters
for both boron (B) and nitrogen (N) (neglecting the B-N interaction) to
investigate the structural properties. Matsunaga
\emph{et al}~\cite{Matsunaga} proposed the TP of B in order to
simulate cubic boron carbonitrides which are compatible with the
parameters of nitrogen fitted by Kroll~\cite{Kroll}, and recently,
Liao \emph{et al}~\cite{Liao} and Sevik~\emph{et al} \cite{Sevik}
reported TP parameters that were used to study the mechanical properties and
the thermal conductivity of a h-BN sheet, respectively.

\begin{figure*}[tp]
\includegraphics[width=0.33\textwidth]{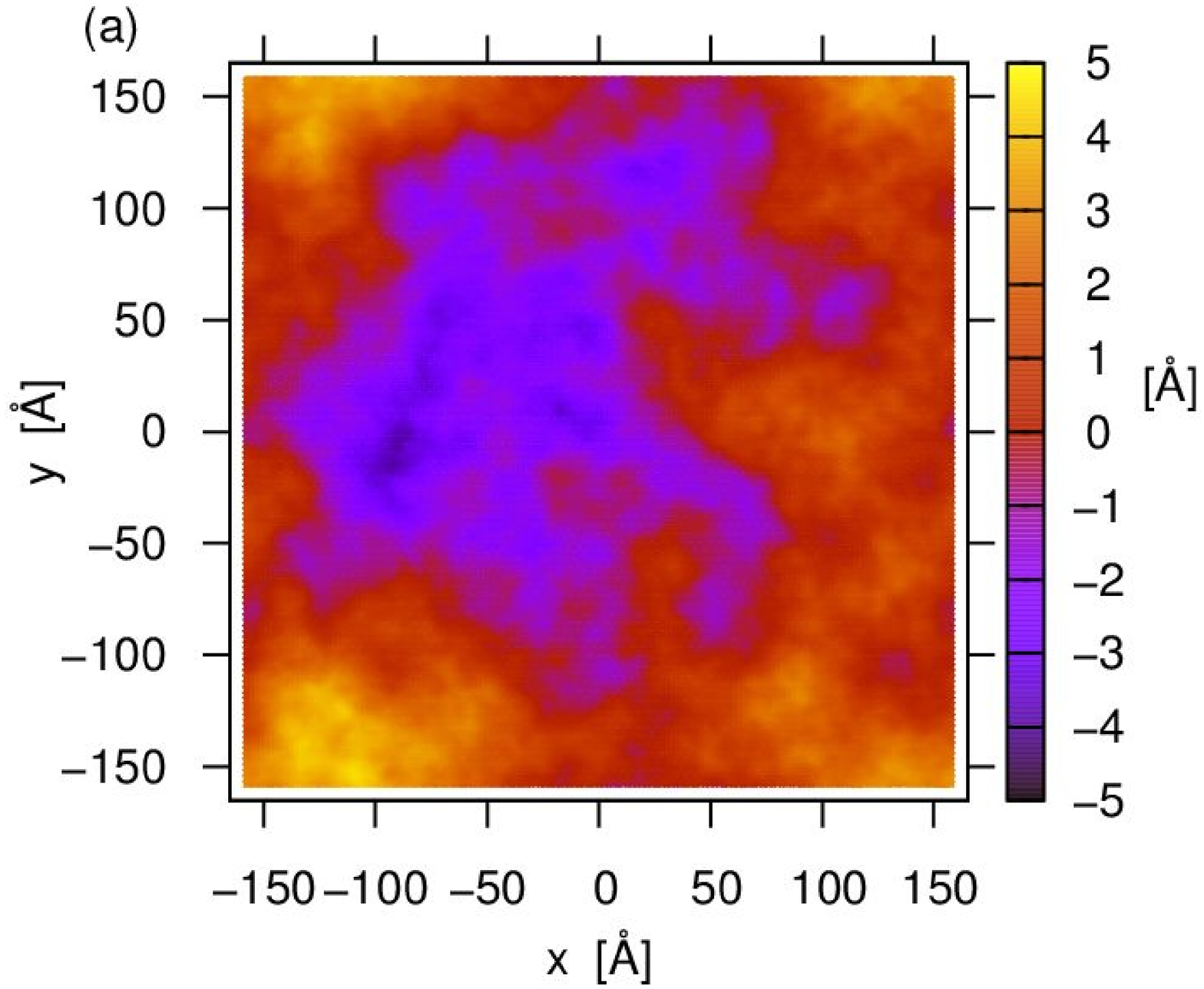}
\hspace{0.1cm}
\includegraphics[width=0.26\textwidth]{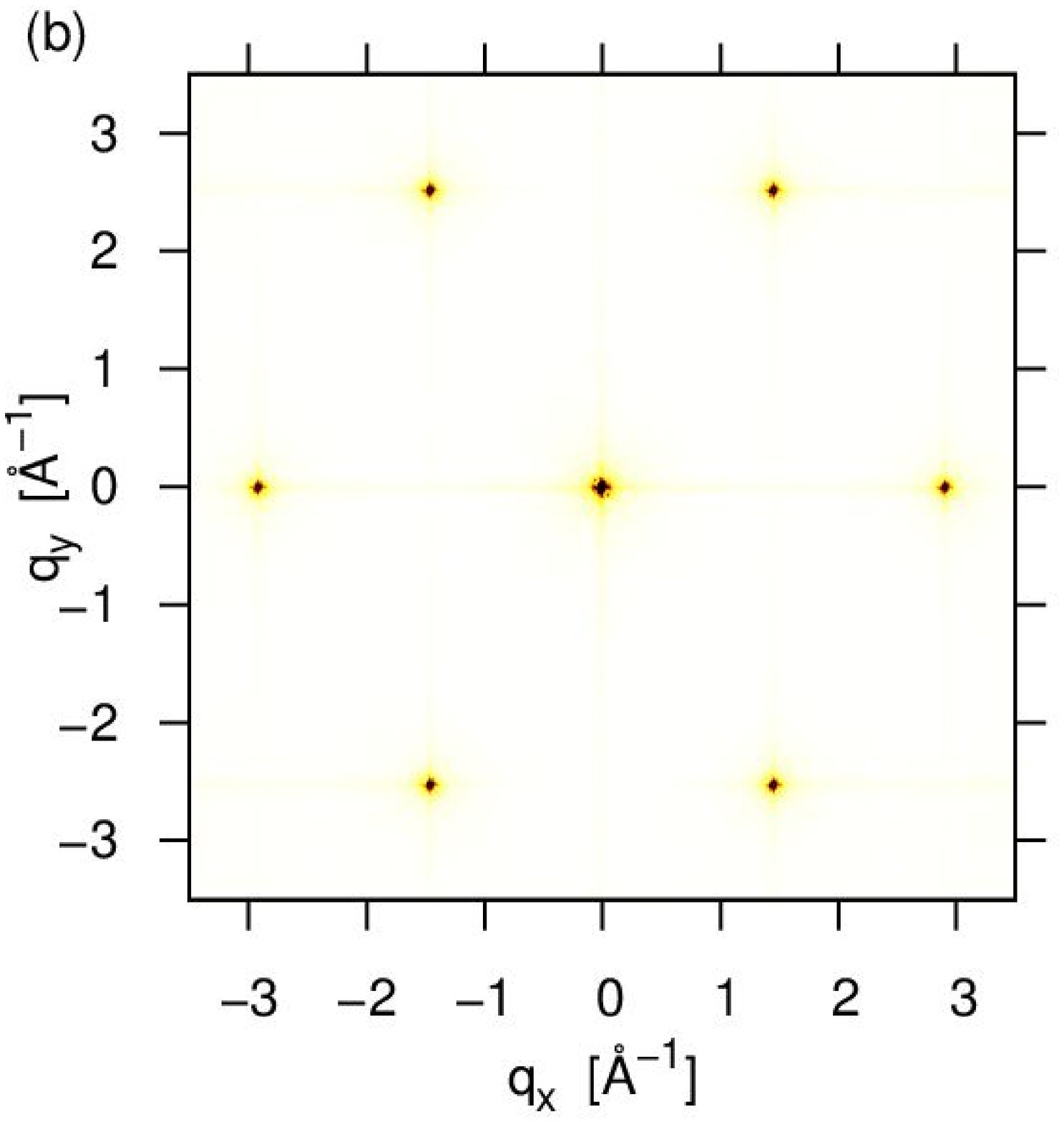}
\hspace{0.1cm}
\includegraphics[width=0.33\textwidth]{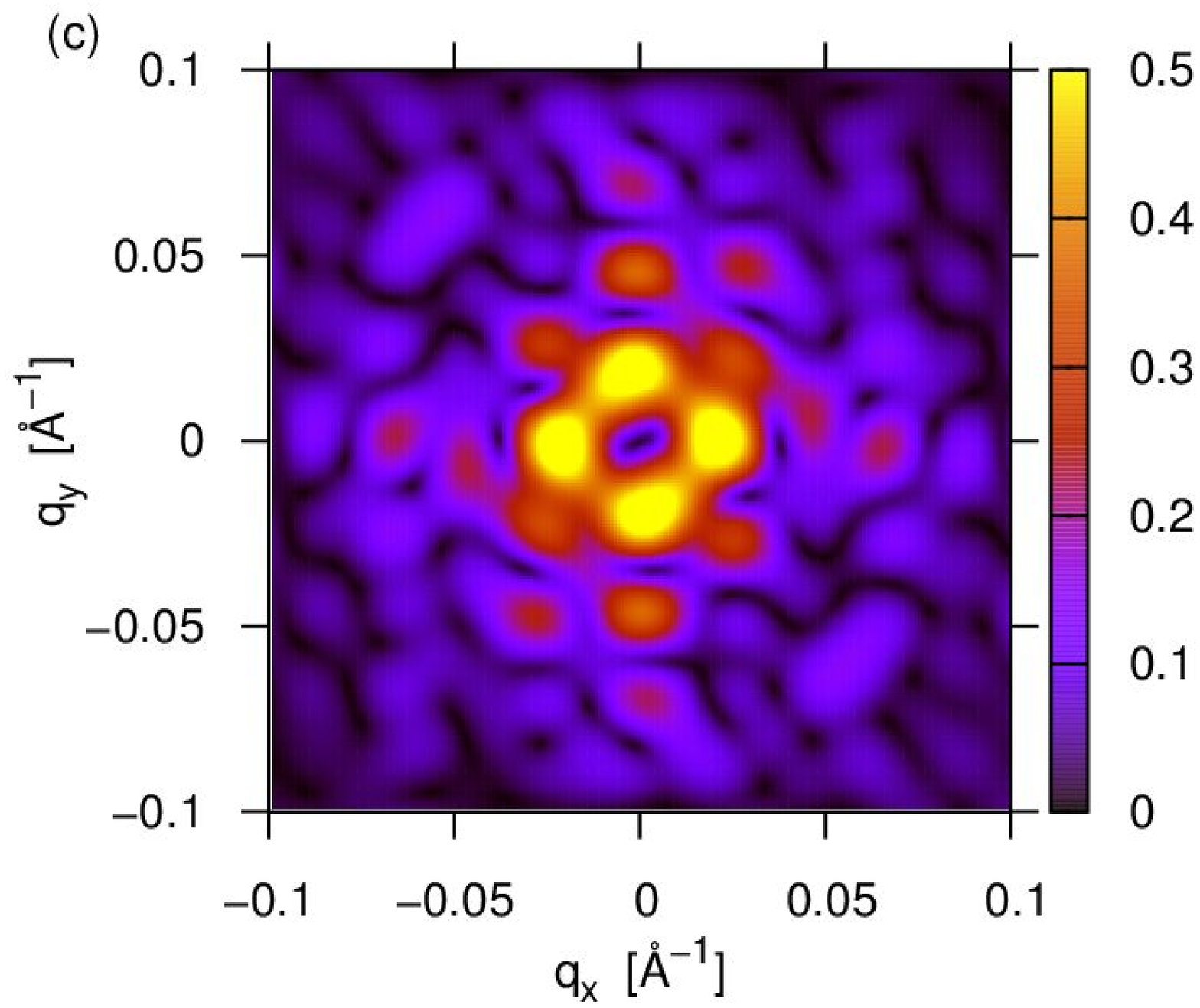}
\caption{ (Color online) (a) Contour plot of the heights for an
arbitrary snapshot taken during the MD simulation at 300\,K. (b)
Corresponding simulated diffraction pattern for the sample shown in
(a). (c) Zoom view of (b) around $\bf{q}$=(0,0).} \label{fig2}
\end{figure*}

In this study, we investigate the thermal rippling behavior of free
standing monolayer h-BN  by using state of the art molecular
dynamics (MD) simulations. 
We adopted the TP potential re-parametrized by Sevik \emph{et al}
which has been shown to represent the experimental structure and the
phonon dispersion of the two-dimensional h-BN sheet. We found that
h-BN is more corrugated and a less stiff material as compared to GE.
The height-height correlations can be explained by the theory for
continuum membranes~\cite{NelsonPiranWeinberg}. In addition, we
report results of both uniaxial and shear stress of a h-BN sheet and
compare it with those found for GE.
The buckling transition for compressed h-BN occurs earlier than for
GE and the induced pattern of ripples when subjected to either
uniaxial or shear stress shows significant differences.

This paper is organized as follows. In Sec. II, we introduce the
atomistic model and the simulation method. Then, in Sec. III we
analyze the behavior of the thermal ripples of a h-BN sheet. Here,
we obtain the bending rigidity, the heat capacity, the thermal
expansion coefficients, and we study the buckling transition of
the h-BN sheet when uniaxial external strain and shear
stress are applied. All the results are compared with the ones obtained for a
GE sheet. Finally, in Sec. IV, we present our conclusions.

\section{Methods}

Classical atomistic molecular dynamics (MD) simulation is employed
to simulate large scale samples of h-BN and GE at temperatures
varying from 10\,K up to 1000\,K. We used a modified Tersoff
potential (which is an ordinary defined potential in the LAMMPS
package~\cite{lammps,Plimpton}) according to the parameters proposed
by Sevik~\emph{et al} for the h-BN sheet. All the parameters and a
detailed description of the potential energy is given in
Ref.~[\onlinecite{Sevik}].
To simulate GE we have used the AIREBO potential~\cite{AIREBO} which
is suitable for simulating hydrocarbons. We employed NPT ensemble
simulations with $P$=0 using the Nos\'{e}-Hoover thermostat which
enables us to allow the size of the system to equilibrate (i.e. the
size of the system is not fixed).
All the reported values have been computed averaging over $300-400$
snapshots taken over uncorrelated samples.

We start with a square shaped h-BN sheet with periodic boundary conditions and initial dimensions
322$\AA$~$\times$321$\AA$ (315$\AA$~$\times$315$\AA$~for GE) in the
$x$ and $y$ direction, respectively, which correspond to a total
number of $N$=37888 atoms,
and which are sufficiently large in order to
capture the long-wavelength regime. Periodic boundary
conditions were adopted in both directions.

To analyze the thermal ripples we computed the diffraction
pattern which is typically studied in experiments to detect the
size and shape of the corrugations~\cite{3}. We obtained also
the mean square value of the out-of plane displacements $\langle h^2
\rangle$ of the atoms and, by following previous
works~\cite{fass2,seba}, the height-height correlation function
$\langle H(q) \rangle$ which was determined by considering an average of
the height between the first neighbors of each atom.
The latter was shown to follow a
$q^{-4}$ behavior that is expected from the theory of continuum
two-dimensional membranes at large values of $q$ in the harmonic
approximation (see below).
To analyze the differences between the strain induced corrugations
in the h-BN and the GE sheets we applied uniaxial and shear stress
on both systems as is schematically shown in Fig.~\ref{fig1}. In
order to apply strain we set the force on the two ends equal to zero
and move the end atoms with an infinitesimal compression step
($\delta x=0.01\AA$) in the desired direction. After each
compression step we wait for 2\,ps to allow the system to relax and
to ensure that the system is in thermal equilibrium~\cite{mehdi1}.
Uniaxial compressive stress is applied along the zig-zag or armchair
direction separately, and the shear stress is applied on the
armchair edges with the opposite velocity for the top and bottom
edges. The details of the used  method of applying the boundary
stress can be found in our previous
studies~\cite{mehdi1,mehdi2,mehdi22}

The TP function~\cite{Tersoff} used in the LAMMPS
package~\cite{lammps,Plimpton} can be expressed as

\begin{equation}
{E}=\sum_{i} E_{i}=\frac{1}{2}\sum_{i\neq j)} \phi(r_{ij}),
\end{equation}
with
\begin{equation}
\phi(r_{ij})=\sum_{i} \sum_{j(>i)} f_{c}(r_{ij})[f_R(r_{ij})+b_{ij}f_A(r_{ij})],
\end{equation}

\noindent where  $f_{c}$, $f_{R}$ and $f_{A}$ are cutoff
functions, the repulsive pair term, and the attractive pair term,
respectively. $r_{ij}$ and $b_{ij}$ are respectively the distance from atom $i$
to atom $j$ and the bond order function.
The use of TP disregard contributions coming from charge
re-distribution which may occur in an ionic crystal. The inclusion
of this effect in h-BN modifies the phonon spectrum significantly only for
energies corresponding to the optical modes~\cite{michel1,michel2}.
The large scale thermal ripples addressed here are governed mainly
by the transversal acoustic mode (TA), which accounts for out-of
plane fluctuations, and it couples with the in-plane modes which
renormalizes the long wave-length behavior, e.g. the bending
rigidity $\kappa$ can be calculated directly from the TA
mode~\cite{fas-phon}.
Therefore, the charge redistribution is expected not to affect the thermal fluctuations analyzed
here and the use of the TP is justified~\cite{reax}.
%

\section{Results and Discussion}
\subsection{Thermal excited ripples}

\begin{figure}[t]
\vspace{0.5cm}
\includegraphics[width=0.485\textwidth]{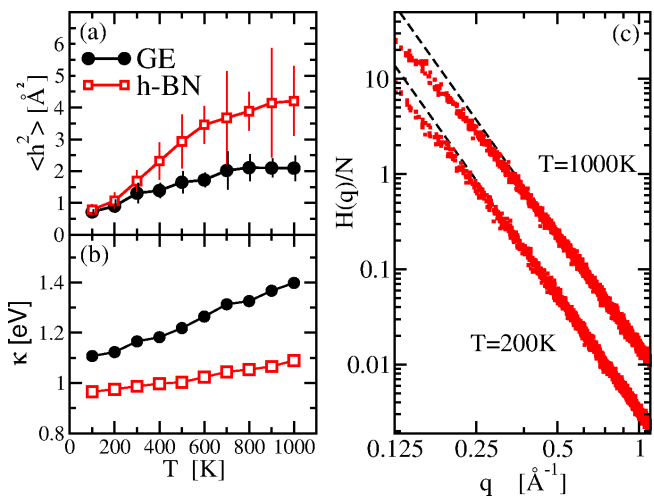}
\caption{(Color online) Variation of (a) $<h^2>$  and (b) the bending rigidity
versus temperature for both h-BN (open squares) and GE (solid circles).
(c) Height-height correlation function $H(q)$ of h-BN at two different temperatures
as is indicated. The dashed lines are the harmonic $q^{-4}$ law.}
\label{fig3}
\end{figure}

%

In Fig.~\ref{fig2}(a) we show a height contour plot of the atoms of
the h-BN sheet for an arbitrary snapshot taken during the MD
simulation at 300 K. The corresponding modeled diffraction pattern
is shown in Fig.~\ref{fig2}(b). This pattern is very similar to the
one obtained for the GE sheet~\cite{3} with the main difference in
the distance between the Bragg points due to the unequal lattice constant of h-BN and GE.
From the zoom plot around $\bf{q}$=(0,0) (Fig.~\ref{fig2}(c))  one
observes the local structure of the central Bragg point
for these intrinsic thermal ripples of h-BN. Notice that the lack of the
presence of the $\bf{q}$=(0,0) component is a consequence of the absence of a perfectly flat h-BN sheet.

The signatures of the thermal induced ripples can also be seen in
the mean square value of the out-of-plane fluctuations $\langle
h^2\rangle$. In Fig.~\ref{fig3}(a) we show $\langle h^{2} \rangle$
as function of temperature. In comparison with GE
(included also in this figure for comparative purposes) we observe that
$\langle h^{2} \rangle$ is always larger for the h-BN sheet.
The weaker (stronger) the atomic B-N (C-C) bonds, the larger (smaller)
the corrugations in h-BN (GE). Notice that the B-N bond is not pure
covalent and it has an ionic character which is due to the
different electronegativity between the two elements, i.e. the electrons
are localized closer to the N atoms rather than the B atoms.
Although this partially ionic structure of the h-BN layer increases
the interlayer interaction resulting in a larger hardness of 3D bulk
h-BN relative to graphite, it makes the single layer of BN  less
stiff than graphene. Moreover, the stacking of h-BN sheets in bulk
h-BN is AAA stacking which is different for the ABC stacking in
graphite~\cite{PRB802009}.
%

\subsection{Bending rigidity {\large {\bf $\kappa$}}}

Accordingly to the two-dimensional theory of continuum membranes the height-height correlation function,
in the harmonic approximation, where the out-of-plane and in-plane modes are decoupled,
is expected to behave as
\begin{eqnarray}
H(q)=\langle|h(q)|^2\rangle=\frac{N k_B T}{\kappa S_0 q^{4}},
\label{hq1}
\end{eqnarray}
where $\kappa$ is the bending rigidity of the membrane,
$N$ is the number of atoms of the sample, $S_0$ the surface
area per atom and $k_B$ is the Boltzmann constant.
In the large wavelength limit, i.e. for $ q \to 0$,  the height fluctuations
are suppressed by anharmonic couplings between bending and
stretching modes giving rise to a renormalized $q$-dependent bending
rigidity and hence Eq.~(\ref{hq1}) is no longer valid~\cite{ledousal,fass2,seba}.
In Fig.~\ref{fig3}(b) we show $\kappa$ for h-BN calculated from the
harmonic part of $H(q)$ between $q$=0.5 $\AA^{-1}$ and $q$=1
$\AA^{-1}$. In agreement with the larger values of $\langle h^2
\rangle$, we observe that the h-BN membrane posses a smaller
$\kappa$ as compared to GE and in the whole temperature range it is about
$16~\%$ smaller at room temperature (300\,K).
Note that $\kappa$ for both h-BN and GE increase with temperature.
In Fig.~\ref{fig3}(c) we show $H(q)$ at 200\,K and 1000\,K were the harmonic behavior can
be clearly observed (fitted with a dashed line) and, as expected,
with increasing temperature $H(q)$ is shifted to larger $q$.
This figure also reveals that the ripples are not characterized by
an unique wave-length and instead follows the behavior expected from
continuum membrane theory.

\textbf{Before ending this section it is worthwhile to investigate
an alternative method to estimate the bending rigidity (flexural
rigidity). A common method for calculating the bending rigidity of
BN-sheet is by performing simulations of BN-nanotubes as a
function of its radius (R) of the curved tubes and then extrapolating the
results to R$\rightarrow\infty$. Hence, one can calculate the
elastic energy of the nanotube as a function of the inverse square of
the radius, $E= \frac{1}{2}\kappa R^{-2}$. This method was used in
Ref.~[25] and in our previous work~\cite{mehdi22} to calculate the
zero temperature bending rigidity of GE and h-BN which were
found to be 1.46\,eV and 1.29\,eV, respectively. The result obtained with the
Tersoff potential using Eq.~(3) is less than the result of Ref.
[25]. The bending rigidity of GE is
larger than h-BN  with about 0.15\,eV in agreement with Ref.[25].
In order to have an independent check we estimated the bending rigidity of a BN sheet by
performing several ground state simulations for (n,n) BN-nanotubes
with n=5,.., 20 using the Tersoff potential. Figure 4 shows the variation of the strain
energy per atom as function of the inverse square radius of the tube.
Fitting a line to the data and dividing by the area of half of unit cell atom results
in $\kappa=0.86$\,eV. The result of Tersoff potential agrees well
with our estimation for $\kappa$ using Eq.~(3) (Ref.~[25]). Thus we can conclude that result based on
the Tersoff potential underestimate the bending rigidity as
compared to the result from  DFT.}

\begin{figure}[t]
\vspace{0.5cm}
\includegraphics[width=0.485\textwidth]{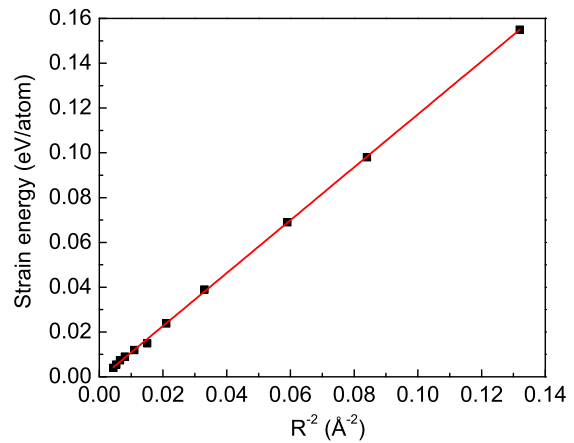}
\caption{(Color online) \textbf{Variation of strain energy versus
inverse square of BN-nanotube radius using Tersoff potential.}} \label{fig34}
\end{figure}

\subsection{Heat capacity, thermal expansion and Gruneisen parameter}
\begin{figure}[t]
\includegraphics[width=0.45\textwidth]{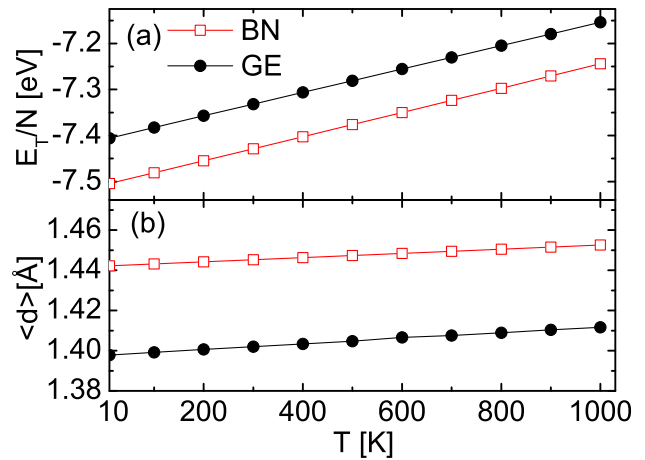}
\caption{(Color online) Variation of (a) the total energy per atom and
(b) the averaged bond length versus temperature for h-BN and GE
(The error bar is less than $10^{-4}$ eV/atom for the total energy).
}\label{fig4}
\end{figure}

The variation of the total energy per atom with temperature
($\geq10K$) is shown in Fig.~\ref{fig4}(a). The total energy varies
linearly with temperature and gives the corresponding lattice
contribution to the molar heat capacity at constant pressure (the
average size of the system after relaxation is taken constant) which
for the h-BN sheet results into
\begin{equation}
C_{P}=\frac{dE_{T}}{dT}=25.2(3)\,J\,mol^{-1}K^{-1},
\end{equation}
\noindent which is comparable to the one for a GE sheet, i.e. 24.5(9)\
$J\,mol^{-1}K^{-1}$ and close to the well known classical molar heat
capacity, i.e., $C=3\Re\simeq24.94$ $J\,mol^{-1}K^{-1}$, i.e., the
Dulong-Petit value, where $\Re$ is the universal gas constant.
\textbf{ Notice that the  heat capacity is a temperature dependent
parameter that will decrease for temperature below the Debye temperature. However, the
classical MD simulation gives the  correct high temperature asymptotic limit
i.e., the Dulong-Petit value, but fails in the low and intermediate temperature range.}

In Fig.~\ref{fig4}(b) we show the variation of the averaged B-N bond length with temperature.
The linear behavior enables us to
calculate the linear thermal expansion of the lattice parameter of a
h-BN sheet as
\begin{equation}
\gamma_{BN}=\frac{1}{a}\frac{d a}{d T} =7.2(1) \times 10^{-6} K^{-1}
\end{equation}
\noindent where $a$=1.442\,\AA~is the zero temperature lattice
parameter. The obtained $\gamma$ is 33$\%$ larger than the one
measured for cubic BN, i.e. 4.8$\times 10^{-6}
K^{-1}$~\cite{diamond} and is comparable to the one found for GE,
i.e. $\gamma_{GE}=10.0(7)\times10^{-6}\,K^{-1}$. The latter (i.e.
$\gamma_{GE}$) is in good agreement with our previous
studies~\cite{mehdi1,mehdi2}.

From  the obtained values of $\gamma$ and $C$ we can compute the Gruneisen parameter
\begin{equation}
\alpha_{BN}=\frac{\gamma B}{C \rho}=0.89,
\end{equation}
\noindent where $B$ is the two dimensional bulk modulus and $\rho$
is the mass density. Using $B_{h-BN}$=3\,eV\AA$^{-2}$ ($B_{GE}=12.7$
eV\AA$^{-2}$ [\onlinecite{fasolino}])),
$\rho_{h-BN}=24.82/S_{h-BN}=3.81\times10^{-4}$g\,m$^{-2}$
($\rho_{GE}=7.6\times10^{-4}$g\,m$^{-2}$)  for h-BN (GE), we obtain
$\alpha_{BN}=0.89$ ($\alpha_{GE}$=1.2). Note that the estimated
value of $\alpha_{GE}$ is better than the one found in
Ref.~[\onlinecite{Tersoff2009}], i.e. -0.2, and is closer to the
experimental result, i.e. 2.0~[\onlinecite{Gruneisen}].


\subsection{Buckling transition}

\begin{figure}
\begin{center}
\includegraphics[width=0.45\textwidth]{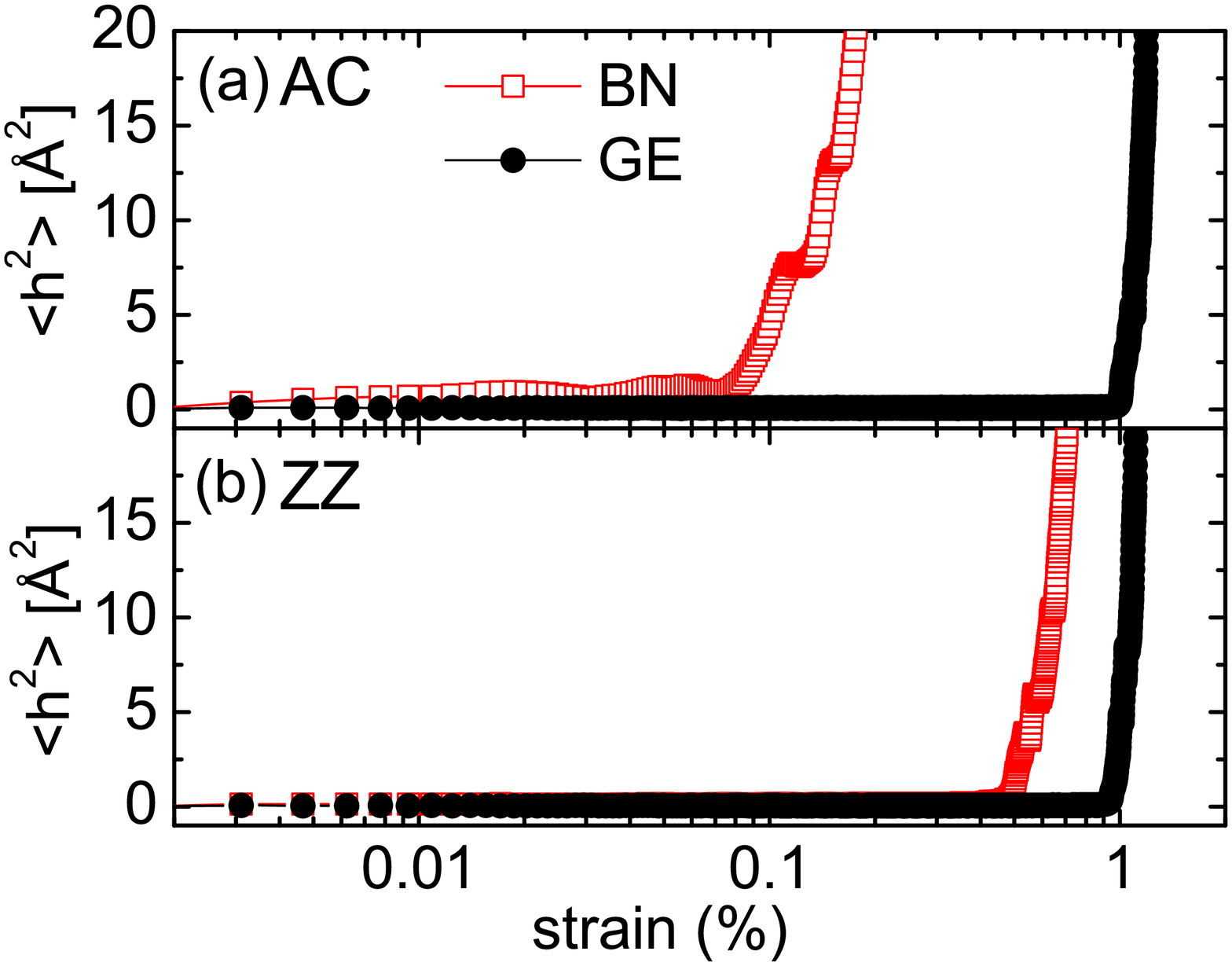}
\includegraphics[width=0.8\linewidth]{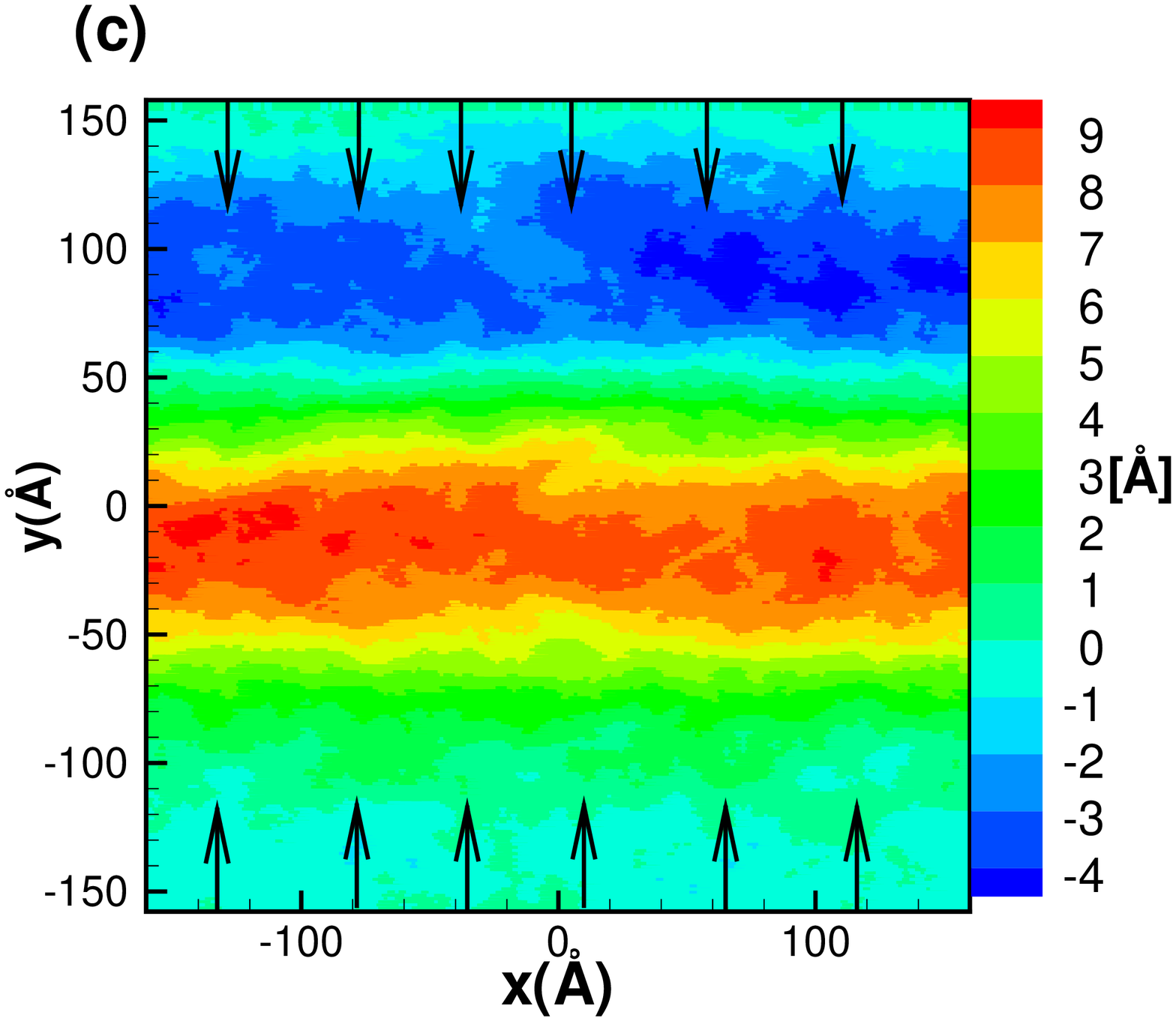}
\hspace{0.08cm}
 \caption{ (Color online)
Variation of $<h^2>$ versus external uniaxial strain in the (a)
armchair and (b) zig-zag direction. Contour plot of the heights of
compressed h-BN sheet samples subjected to a compressive strain for
fixed $<h^2>= 20$ $\AA^{2}$ for compression in (c) zig-zag direction
of the h-BN sheet. Arrows indicate the stress
direction.}\label{fig5}
\end{center}
\end{figure}
\label{strain}

\begin{figure}
\begin{center}
\includegraphics[width=0.45\textwidth]{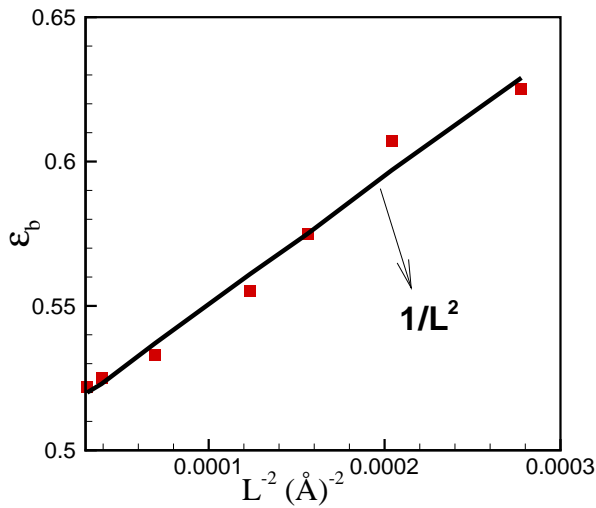}
 \caption{ (Color online) Variation of buckling strain with the longitudinal length
 of the BN sheet which is compressed along the zig-zag direction.}\label{figEuler}
\end{center}
\end{figure}

The different stiffness between the h-BN and GE sheets results in
different buckling transitions. To study this, we compressed
the systems uniaxially where we considered  both zig-zag and armchair directions. This
was done by fixing one row of atoms in each edge during the
compression steps as is indicated schematically in Fig.~\ref{fig1}.
The compression rate was taken $\mu=0.5 m/s$ which is similar to the one used
in our previous study~\cite{mehdi2}, and small enough to guarantee
that the system is in equilibrium during the whole compression
process~\cite{PRLneek}. The simulations were carried out at room
temperature. 
Figure~\ref{fig5}(a) shows the variation of $<h^2>$ versus strain,
i.e. $\epsilon=\mu t/l$ where $t$ is the time (after starting the
compression) and $l$ is the initial box length in the direction of
the compression. 
The buckling transition for the h-BN sheet kept at room temperature
is found to be $0.6$ ($0.1$) which is smaller than the one for GE,
i.e. $1.0$ ($1.2$), for uniaxial compression along the zig-zag
(armchair) direction~\cite{suppl}.
Hence, when the compression is applied in the zig-zag direction, the
h-BN sheet resists much more against the external uniaxial stress as
compared to the case the stress is applied along the armchair
direction.
\textbf{The smaller critical strain at which the  buckling transition in the armchair direction of
the h-BN sheet takes place indicates its more sensitive nature to external
uniaxial stress~\cite{compmat2012}.} Although the same argument
holds for GE, the difference between the two directions is much
smaller. \textbf{Notice that  DFT calculations result in a
deviation in bending rigidity (flexural rigidity) between ZZ BN-nanotube
and AC BN-nanotube (ZZ becomes larger than AC) for radius
smaller than $\simeq 3\AA~[25]$ while for larger radius they are the
same.}

 A contour plot of the buckled h-BN sheet with $\langle h^2
\rangle$=20 $\AA^2$ and compressed in the zig-zag direction is shown
in Fig.~\ref{fig5}(c). The obtained buckling transition for GE, i.e.
0.8$\%$, agrees very well with our previous studies~\cite{mehdi1} and
is in the range of the experimental measured buckling transition for
suspended GE, i.e. 0.5$\%$-0.7$\%$~\cite{8,9}.

It is also interesting to compare the buckling strain with those
predicted by the theory of loaded beam. Euler theory for a two
end-loaded beam having length $L$ predicts that $\epsilon_b\propto
L^{-2}$~\cite{buckling,mehdi2}. \textbf{The first demonstration by
MD of Euler buckling in nanostructures goes back to the pioneer work
by Yakobson~\emph{et al}~\cite{Yakobson1996}}. We performed several
simulations in order to find the length dependence of the buckling
strain for BN sheets which are e.g. compressed along the zig-zag
direction. Figure~\ref{figEuler} shows the variation of $\epsilon_b$
with the longitudinal length ($L$) of the zig-zag BN sheet. The
solid line is a fit to $L^{-2}$ and the symbols are the results of
our MD simulations which are qualitatively in agreement with
 the theory of loaded beam.

\begin{figure*}
\begin{center}
\hspace{0.08cm}
\includegraphics[width=0.46\textwidth]{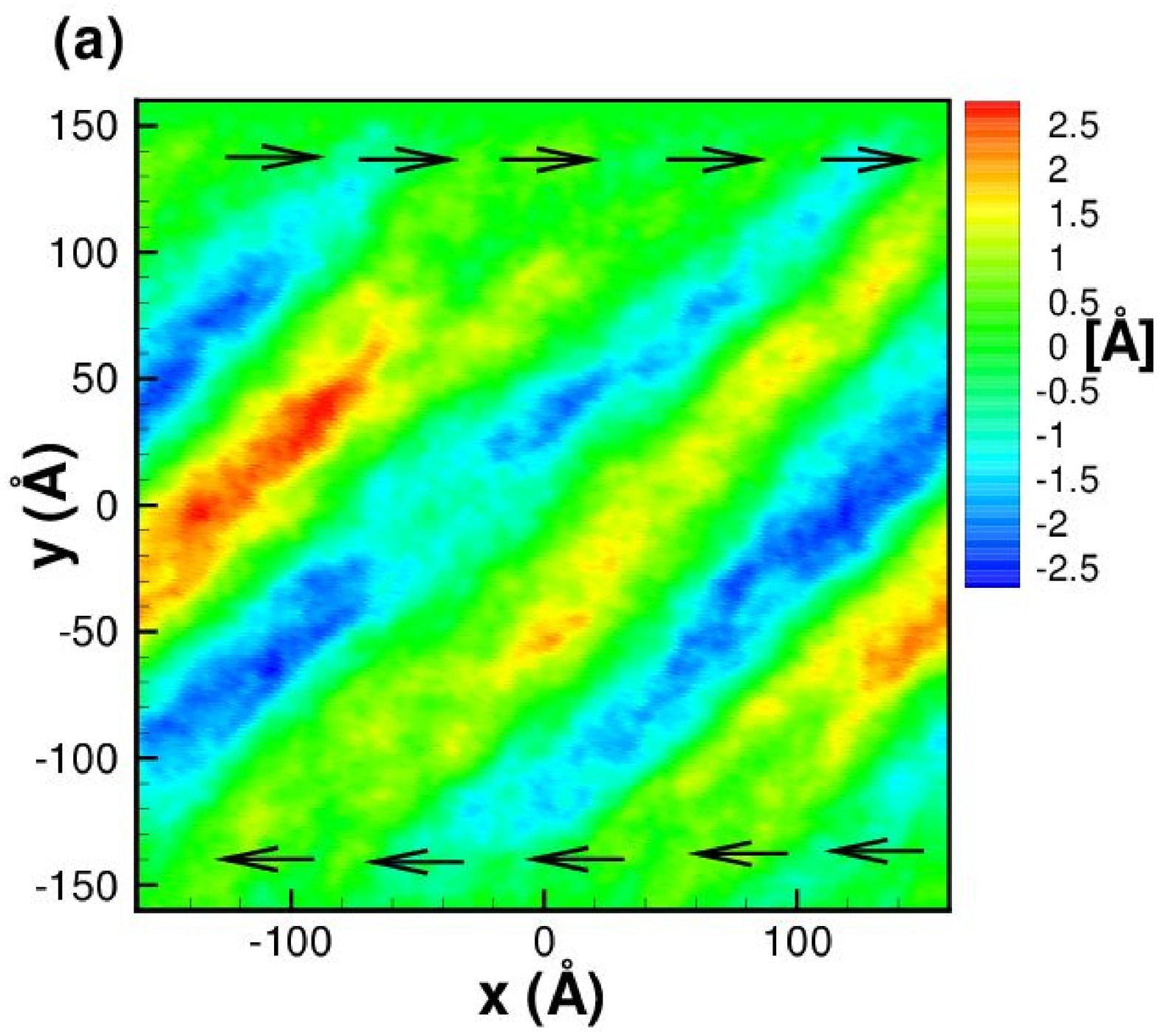}
\includegraphics[width=0.46\textwidth]{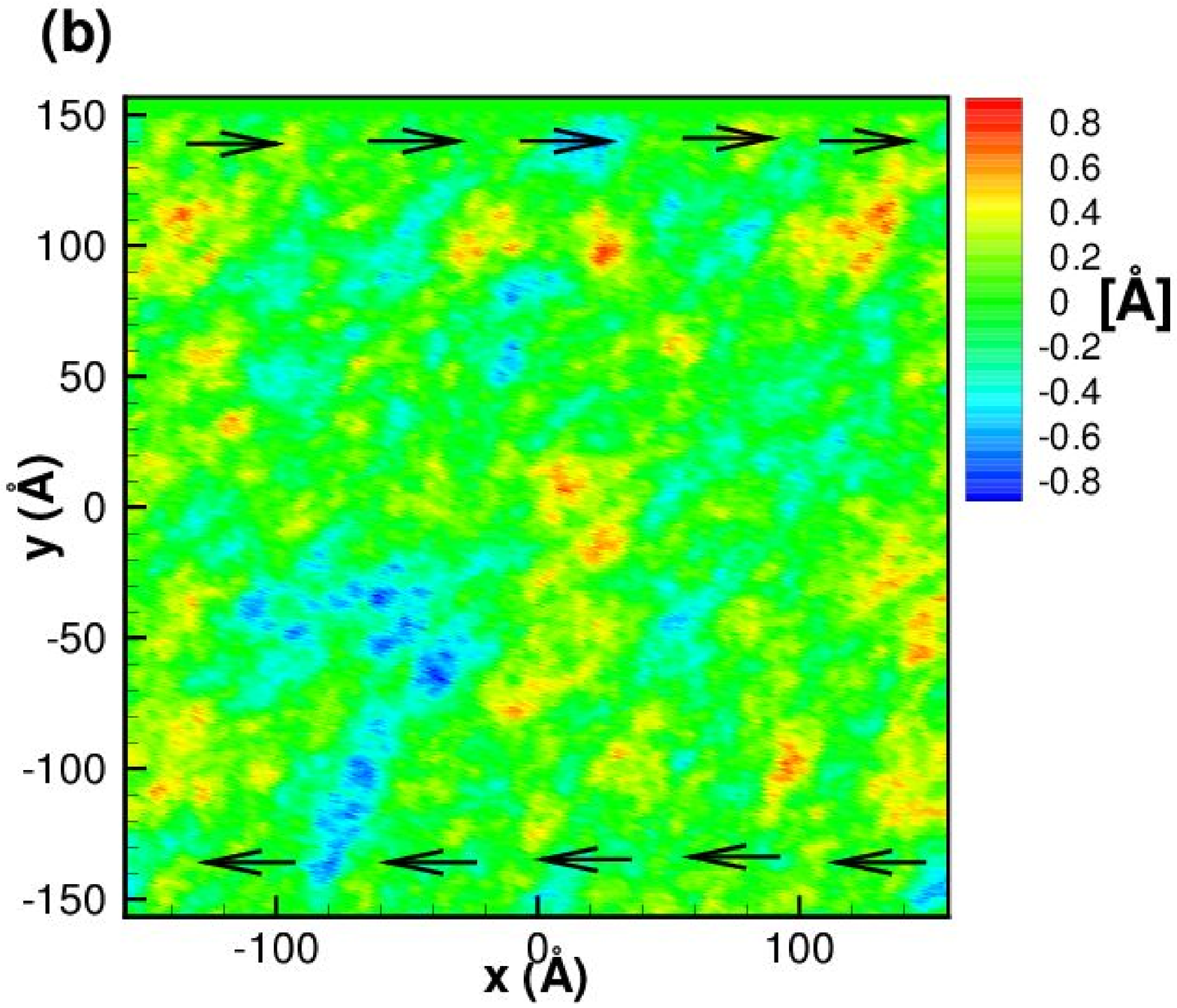}
\hspace{0.08cm}
\caption{Effect of a shear stress of about 1.5 $\%$ on the out-of-plane deformation of (a) the h-BN sheet
and (b) the graphene sheet. Arrows indicate the stress direction on the armchair edges.
Note the larger values of the heights in h-BN.}
\label{fig6}
\end{center}
\end{figure*}

Finally, we report the results  for h-BN
and GE sheets under shear stress. Here we applied the stress only in the armchair
direction as is described schematically in Fig.~\ref{fig1}. We found
that, due to the larger stiffness, to reach the same value of
$\langle h^2\rangle$ a larger shear stress has to be applied in
GE as compared to h-BN. In Figs.~\ref{fig6}(a,b) one can observe
significant differences between the corrugated configurations of
h-BN and GE sheets. These samples were subjected to a shear stress
of $\epsilon=1.5\%$. While h-BN is highly sensitive to  shear
stress and deforms easily, GE can resist larger stress values due to
its larger rigidity.

\begin{section}{Conclusions}

The thermal properties of a boron nitride sheet were studied using large
scale atomistic simulations. We showed that the scaling properties of a h-BN sheet
follows closely the results of membrane theory and hence the thermal excited ripples
are not characterized by any particular wave-length. Using the harmonic part of
the height-height correlation function we found an increasing bending rigidity
with temperature which is smaller than the one of graphene.
We found that the buckling transition for h-BN depends on the applied
compression direction and is much smaller than the one of graphene.
The obtained molar heat capacity agrees very well with the
well-known Dulong-Petit number, 25.2 $\ J\ mol^{-1} K^{-1}$ and the
thermal expansion coefficient was found to be positive and equal to
7.2 $\times 10^{-6} K^{-1}$. The Gruneisen parameter 0.89 is found
to be smaller than the one for graphene, i.e. 1.2.  We showed that
the different stiffness between the GE and h-BN sheets leads to
different patterns of deformations in the presence of either
uniaxial or shear stress.

\end{section}
\vspace{0.25cm}

\vspace{0.25cm}
\section*{Acknowledgments}
We thank K. H. Michel and D. A. Kirilenko for their useful comments
on the manuscript. M. N.-A. is supported by the EU-Marie Curie IIF
postdoc Fellowship/299855. S. Costamagna is supported by the Belgian
Science Foundation (BELSPO). This work is supported by the
ESF-EuroGRAPHENE project CONGRAN, the Flemish Science Foundation
(FWO-Vl), and the Methusalem programme of the Flemish Government.

\end{document}